\documentclass[aps,prl,reprint,twocolumn,superscriptaddress,longbibliography,nofootinbib,floatfix,showpacs]{revtex4-1}
\usepackage{epsfig}
\usepackage{amsmath,amssymb,amsfonts}
\usepackage{mathrsfs}
\usepackage{bbm}
\usepackage{slashed}
\usepackage{graphicx}
\usepackage{verbatim}
\usepackage[T1]{fontenc}
\usepackage[utf8]{inputenc}
\usepackage[colorlinks]{hyperref}
\usepackage{ifthen}
\usepackage{forloop}
\usepackage[english]{babel}
\usepackage{mathbbol}
\usepackage[usenames]{color}

\definecolor{darkgreen}{rgb}{0.2,0.6,0}
\definecolor{lightblue}{rgb}{0,0.5,0.8}
\definecolor{lightred}{rgb}{0.8,0.2,0.2}
\definecolor{darkorange}{rgb}{1,0.549,0}

\newcommand{\cF}{{\mathcal F}}

\newcommand{\UV}{{\small UV}}
\newcommand{\IR}{{\small IR}}

\newcommand{\CDT}{{\small CDT}}
\newcommand{\ADM}{{\small ADM}}
\newcommand{\eg}{{\textit{e.g.}}}

\graphicspath{{figures/}}

\newcommand{\be}{\begin{equation}}
\newcommand{\ee}{\end{equation}}
\newcommand{\ba}{\begin{eqnarray}}
\newcommand{\ea}{\end{eqnarray}}

\newcommand{\cR}{{\mathcal R}}

\begin{document}

\title{Towards reconstructing the quantum effective action of gravity}
\author{Benjamin Knorr}
\email[]{b.knorr@science.ru.nl}
\author{Frank Saueressig}
\email[]{f.saueressig@science.ru.nl}
\affiliation{
Institute for Mathematics, Astrophysics and Particle Physics (IMAPP),\\
Radboud University Nijmegen, Heyendaalseweg 135, 6525 AJ Nijmegen, The Netherlands
}

\begin{abstract}
Starting from a parameterisation of the quantum effective action for gravity we calculate correlation functions for observable quantities. The resulting templates allow to reverse-engineer the couplings describing the effective dynamics from the correlation functions. Applying this new formalism to the autocorrelation function of spatial volume fluctuations measured within the Causal Dynamical Triangulations program suggests that the corresponding quantum effective action consists of the Einstein-Hilbert action supplemented by a non-local interaction term. We expect that our matching-template formalism can be adapted to a wide range of quantum gravity programs allowing to bridge the gap between the fundamental formulation and observable low-energy physics.
\end{abstract}

\maketitle
\section{Introduction}
A characteristic feature of quantum gravity research is its fragmentation into disjoint branches including, \eg{}, string theory \cite{Becker:2007zj, Schomerus:2017lqg}, loop quantum gravity \cite{Thiemann:2002nj, Rovelli:2004tv}, the Asymptotic Safety program \cite{Niedermaier:2006wt, Reuter:2012id, Percacci:2017fkn, Eichhorn:2017egq}, Causal Dynamical Triangulations (\CDT{}) \cite{Ambjorn:2012jv}, Causal Set Theory \cite{Sorkin:2003bx, Surya:2011yh}, Group Field Theory \cite{Baratin:2010wi, Oriti:2011jm} or non-local gravity theories \cite{Biswas:2005qr, Modesto:2011kw, Modesto:2017sdr}. Each approach formulates its own fundamental dynamics typically at the Planck scale. The complexity of these formulations makes it quite hard to derive physical consequences from the fundamental formulation.\footnote{Any candidate for a theory of quantum gravity that reproduces General Relativity in the infrared gives rise to universal 1-loop corrections to \eg{} the Newton potential. It is however highly non-trivial that a given microscopic prescription based, e.g., on a discretized spacetime structure or the violation of Lorentz symmetry at trans-Planckian scales, admits such a limit.}

 A canonical way towards addressing this problem would be the construction of the quantum effective action $\Gamma$ which encodes the dynamics of a quantum theory taking all quantum fluctuations into account. In this way it stores the outcome of a large number of (scattering) processes in an economical way. Generally, finding the exact form of $\Gamma$ is considered equivalent to solving the theory. Not surprisingly calculating the quantum effective action is a hard problem. While fundamental Lagrangians describing our known physical theories are local and often restricted to a small number of interaction terms, $\Gamma$ generically contains all possible interactions permitted by the symmetries of the theory. Furthermore, quantum corrections related to massless particles like the graviton give rise to non-local terms \cite{Donoghue:2017pgk}.
 
In contrast, correlation functions built from fluctuations of physical quantities like volumes of (sub-) manifolds $\Sigma$, $V_n = \int_\Sigma \text{d}^nx \sqrt{g}$, and curvatures are accessible even at the non-perturbative level \cite{Ambjorn:2008wc, Ambjorn:2016fbd, Ambjorn:2017ogo, Ambjorn:2018qbf}. In this work we explicitly demonstrate that this information allows to reconstruct (parts of) the underlying quantum effective action, thereby taking a first explicit step in such a reconstruction program. Starting from the two-point autocorrelation functions of three-volume fluctuations measured in Monte Carlo simulations within \CDT{} \cite{Ambjorn:2008wc, Ambjorn:2016fbd, Ambjorn:2017ogo, Ambjorn:2018qbf} we determine several couplings appearing in $\Gamma$. Our analysis provides first-hand evidence for the presence of non-local terms which could affect the gravitational dynamics at cosmic scales. 
\section{The quantum effective action for gravity}

In the case of gravity, the quantum effective action may be built from the spacetime metric and its curvature tensors. The local part of the quantum effective action can then be organised in terms of a derivative expansion. The lowest order terms coincide with the Einstein-Hilbert action
\begin{equation}\label{Glocal}
\Gamma^{\rm local}  = \frac{1}{16 \pi G_N} \int d^4x \sqrt{g} \left[2 \Lambda - R  \right]  + \ldots
\end{equation}
where $G_N$ and $\Lambda$ are Newton's constant and the cosmological constant. The dots represent terms containing four or more derivatives as $\int d^4x \sqrt{g} R^2$, or the Goroff-Sagnotti counterterm \cite{Goroff:1985sz, Goroff:1985th, Gies:2016con} 
which will not be resolved here. 

{The non-local (NL) part of $\Gamma$ typically contains inverse powers of the Laplacian $\Box \equiv - g^{\mu\nu}D_\mu D_\nu$ acting on curvature tensors.
 At second order in the curvature this leads to terms of the form\footnote{For \UV{} modifications of gravity including these types of form factors see \eg{} \cite{Biswas:2011ar, Edholm:2016hbt}.}
\begin{equation}\label{Gnon-local}
\Gamma^{\rm NL} \propto \int d^4x \sqrt{g} \, \cR \, \cF(\Box) \, \cR \, ,
\end{equation}
where $\cR$ is linear in the curvature tensors and the functions $\mathcal{F}(\Box)$ are known as form factors.\footnote{The form factors $\cF(\Box)$, defined through the matrix elements $\langle x | \cF(\Box) |y\rangle \equiv L(x-y)$ allow to write non-local terms $\int d^4x \sqrt{g(x)} \int d^4y \sqrt{g(y)} R(x) L(x-y) R(y)$ into quasi-local form. Regularity of the non-local terms may require supplementing the operator appearing in the structure functions by non-trivial endomorphism terms built from the curvature \cite{Barvinsky:2012mro}. We do not include them in the expressions \eqref{Gnon-local}, since their resolution depends on correlators which are beyond the scope of this work.} For the purpose of this work, we focus on diffeomorphism-invariant contributions which are quadratic in the curvature tensor and contain two inverse powers of the Laplacian, $\cF(\Box) = \Box^{-2}$. This class contains the two independent terms
\begin{subequations}\label{eq4}
	\begin{equation}\label{4a}
	\Gamma^{\rm NL}_R  = \,  - \frac{b^2}{96 \pi G_N} 
	\int d^4x \sqrt{g} \, R \, \Box^{-2} \, R \, , 
	\end{equation}    
	\begin{equation}\label{4b}
	\Gamma^{\rm NL}_C  = \,  - \frac{\tilde{b}^2}{96 \pi G_N} 
	\int d^4x \sqrt{g} \, C_{\mu\nu\rho\sigma} \, \Box^{-2} \, C^{\mu\nu\rho\sigma} \, .
	\end{equation}
\end{subequations}
Using the Bianchi identity satisfied by the Riemann tensor, $\int d^4x \sqrt{g} \, R^{\mu\nu} \, \Box^{-2} \, R_{\mu\nu}$ can be rewritten in terms of these combinations and higher-order curvature terms and is thus not considered in the present construction. 
The characteristic feature of the non-local terms \eqref{eq4} is that the contribute mass terms to the propagators of scalar (eq.\ \eqref{4a}) and graviton fluctuations (eq.\ \eqref{4b}) when expanded around flat space. Thus eq.\ \eqref{4a} is the only diffeomorphism-invariant combination giving rise to a mass-type contribution to the two-point function in a toroidal background studied below. }
%
The task at hand is then to derive the values of the parameters $G_N, \Lambda, b, \tilde{b}, \ldots$ in terms of the parameters defining the fundamental theory.\footnote{For the implementation of this strategy in effective field theory see \cite{Donoghue:2015nba}.} In general, the latter set will vary from theory to theory. For \CDT{} they are given by the bare Newton's constant $\kappa_0$ and the relative size of spatial and time-like lines encoded in $\Delta$ \cite{Ambjorn:2012jv}.

\section{Derivation of the matching template}
The foliation structure, constituting an elementary building block in the \CDT{} program \cite{Ambjorn:1998xu,Ambjorn:2004qm}, suggests to write $\Gamma$ using the Arnowitt-Deser-Misner (\ADM{})-formalism, reviewed \eg{} in \cite{Gourgoulhon:2007ue}. In this case the spacetime metric $g_{\mu\nu}$ is decomposed into a lapse function, a shift vector and a metric $\sigma_{ij}$ measuring distances on spatial slices $\Sigma$ orthogonal to a normal vector $n_\mu$. Curvature tensors constructed from $g_{\mu\nu}$ can be separated into terms containing the intrinsic and extrinsic curvatures defined with respect to the foliation. {In particular, the volume of the spatial slices,
	\be
	V_3(t) \equiv \int_{\Sigma} d^3x \sqrt{\sigma(t,x)} \, , 
	\ee
	where $\sigma$ denotes the determinant of $\sigma_{ij}$, is invariant with respect to a change of coordinates on the spatial slice and constitutes an observable once the foliation is fixed.
}

\CDT{} simulations have been performed for spatial slices $\bar{\Sigma}$ possessing the topology of a 3-sphere \cite{Ambjorn:2008wc} and recently also for toroidal geometry \cite{Ambjorn:2016fbd, Ambjorn:2017ogo, Ambjorn:2018qbf}. Quite remarkably, the resulting profiles for the expectation value of three-volumes $V_3(\bar{\Sigma},t)$ as a function of the Euclidean time parameter $t \in [0,1)$ agrees with spacetime metrics of the form
\begin{equation}\label{background}
\bar g_{\mu\nu} = \text{diag}(1, a(t)^2 \, \bar \sigma_{ij}(x)) \, ,
\end{equation}
where
\be
\begin{array}{lll}
\textrm{torus:}  \; & a(t) = 1 \, , \quad & \bar{\sigma}_{ij} = \delta_{ij} \, , \\
\textrm{3-sphere:} \; & a(t) = \sin(\pi t) 
\, , \quad  & \bar{\sigma}_{ij}(x) = \bar{\sigma}_{ij}^{S^3}(x) \, , 
\end{array}
\ee
with $\bar{\sigma}_{ij}^{S^3}(x)$ the standard metric of the 3-sphere.
For concreteness, we will focus mainly on the toroidal case and only briefly comment on the analogous analysis for the spherical case. In the former
case the measured volume profile is essentially flat.
Requiring that $\bar\Sigma=S^1\times S^1\times S^1$ is a solution to the equations of motion then fixes $\Lambda = 0$. The higher-derivative and non-local terms in the quantum effective action do not contribute to the dynamics for this case.

Motivated by the existence of a well-defined background geometry, one can then study the autocorrelation of three-volume fluctuations around the background,
\begin{equation}\label{volumecorrelator}
\mathfrak V_2(t^\prime,t) = \langle \delta V_3(t^\prime) \delta V_3(t) \rangle \, .
\end{equation}
Based on the quantum effective action, the fluctuations in the spatial metric are defined in the standard way, setting
\begin{equation}
\sigma_{ij}(t,x) =   \bar \sigma_{ij} + \delta\sigma_{ij}(t,x) \, .
\end{equation}
The fluctuations in the 3-volume can then be found by expanding  $V_3(\Sigma, t) \equiv \int d^3x \sqrt{\sigma}$ in powers of the fluctuations. To leading order,
\begin{equation}\label{volfluct}
 \delta V_3(t) = \frac{1}{2} \int \text{d}^3x \, \sqrt{\bar\sigma}  \bar\sigma^{ij} \delta\sigma_{ij} + \mathcal O(\delta\sigma^2)\, .
\end{equation}
Introducing the fluctuation field $\hat\sigma(t,x) \equiv  \bar\sigma^{ij}\delta\sigma_{ij}(t,x)$, the correlator $\mathfrak V_2$ is given by the integral over the propagator $\mathfrak G_{\hat\sigma\hat\sigma}(t^\prime,x^\prime;t,x) \equiv \langle \, \hat\sigma(t^\prime,x^\prime) \, \hat\sigma(t,x) \, \rangle$
\begin{equation}\label{2cor}
 \mathfrak V_2(t^\prime,t) = \frac{1}{4} \int \text{d}^3x^\prime \sqrt{\bar\sigma} \int \text{d}^3x \sqrt{\bar\sigma} \, \langle \, \hat\sigma(t^\prime,x^\prime) \, \hat\sigma(t,x) \, \rangle \, .
\end{equation}

The computation of the two-point function proceeds by expanding the quantum effective action to second order in $\hat{\sigma}$, $\Gamma^{\rm quad} = \frac{1}{32 \pi G} \int d^4x \sqrt{\bar g} \, \hat{\sigma} \, \Gamma^{(2)} \, \hat{\sigma}$. The propagator $\mathfrak G_{\hat\sigma\hat\sigma}(t^\prime,x^\prime;t,x)$ can then be expressed in terms of the eigenvalue spectrum $\{\bar \lambda_n\}$ and normalised eigenfunctions $\Phi_n(t,x)$ of the differential operator $\Gamma^{(2)}$,
\begin{equation}
\mathfrak G = 16 \pi G_N \, \sum_n \frac{1}{\bar{\lambda}_n} \,  \Phi_n^\ast(t^\prime,x^\prime) \Phi_n(t,x) \, .
\end{equation}
For compact spaces and correlation functions involving fluctuations which are averaged over $\bar \Sigma$ the construction of the two-point function can be simplified by the following observation. On compact spaces the eigenfunctions $\Phi_n(t,x)$ can be expanded in a complete set of orthonormal functions $\psi_k(x)$ defined on $\bar \Sigma$,
\begin{equation}\label{basisexp}
\Phi_n(x,t) = \sum_k  \phi_{n,k}(t) \, \psi_k(x) .
\end{equation}
The spatial integrals appearing in \eqref{2cor} then project the expansion \eqref{basisexp} on the spatially constant mode
$ \psi_0(x) \equiv (V_3(\bar \Sigma))^{-1/2}$. Hence
\begin{equation}\label{eq11}
 \mathfrak V_2(t^\prime,t) = 4\pi G_N \,V_3(\bar\Sigma) \, \sum_n\Big.^\prime \frac{1}{\lambda_n} \phi_n^\ast(t^\prime)\phi_n(t) \, ,
\end{equation}
where $\{\lambda_n\}$ is the eigenvalue spectrum of $\Gamma^{(2)}$ restricted to constant spatial modes. The prime indicates that the zero mode should be excluded since it corresponds to an overall rescaling of the volume.

The next step computes $\Gamma^{(2)}$ by expanding the local and non-local terms in $\Gamma$ given by \eqref{Glocal} and \eqref{Gnon-local} to second order in $\hat \sigma$. Restricting to fluctuations which are constant on $\bar\Sigma$, the result reads
\be
\Gamma^{(2)} = \tfrac{1}{3} \left[ \partial_t^2 - b^2 + \tfrac{1}{2} \Lambda \right] \, . 
\ee
%
{Notably, this expression does not contain $\tilde{b}$, indicating that the correlator \eqref{2cor} does not carry information about the non-local graviton mass term.}

Constructing $\mathfrak V_2(t^\prime,t)$ {on a toroidal background} requires the eigenvalues and eigenfunctions of $\Gamma^{(2)}$, solving\footnote{The mode $\hat \sigma$ comes with a wrong-sign kinetic term. Following the \CDT{} study \cite{Ambjorn:2008wc}, we consider the negative of the corresponding operator in the sequel.}
\begin{equation}
 -\phi_n''(t) + b^2 \phi_n(t) = \lambda_n \phi_n(t) \, .
\end{equation}
The solution is readily given in terms of Fourier modes
\begin{equation}\label{eq:toruseigensol}
 \phi_n(t) = e^{2\pi \mathbf{i} n t} \, , \enspace \lambda_n = (2\pi n)^2 + b^2 \, , \quad n \in \mathbb{Z} \, .
\end{equation}
 Based on the spectrum \eqref{eq:toruseigensol} the propagator can be readily calculated. Carrying out the sum gives
\begin{widetext}
\begin{equation}
 \sum_n\Big.^\prime \frac{1}{\lambda_n} \phi_n^\ast(t^\prime)\phi_n(t) = \frac{e^{2\pi \mathbf{i}(t'-t)}}{2b(4\pi^2+b^2)} \left[ (b-2\pi \mathbf{i}) \, {}_2F_1\left(1,1-\frac{\mathbf{i}b}{2\pi};2-\frac{\mathbf{i}b}{2\pi} \Bigg| e^{2\pi \mathbf{i}(t'-t)}\right) - (b\to-b) \right] + c.c.
\end{equation}
For $b^2 < 4\pi^2$, the sum can be expanded in a convergent power series in $b^2$ whose coefficients contain polylogarithms:
\begin{equation}
 \sum_n\Big.^\prime \frac{1}{\lambda_n} \phi_n^\ast(t^\prime)\phi_n(t) = \sum_{n=1}^\infty \left[ \text{Li}_{2n}\left( e^{2\pi \mathbf{i} (t'-t)}\right) + \text{Li}_{2n}\left( e^{-2\pi \mathbf{i} (t'-t)}\right) \right] \left( \frac{b^2}{4\pi^2} \right)^{n-1} \, .
\end{equation}
\end{widetext}
The limit $b \rightarrow 0$ is thus continuous and finite. The expansion for small time steps $|t^\prime-t| \ll 1$ yields
\be
\sum_n\Big.^\prime \tfrac{1}{\lambda_n} \phi_n^\ast(t^\prime)\phi_n(t) \approx
\frac{b \coth(b/2)-2}{2b^2} - \frac{|t^\prime-t| }{2} + \cdots \, ,
\ee
so that the height of the peak is determined by $b$.

\section{Comparison with CDT data}

\begin{figure}[t]
\includegraphics[width=\columnwidth]{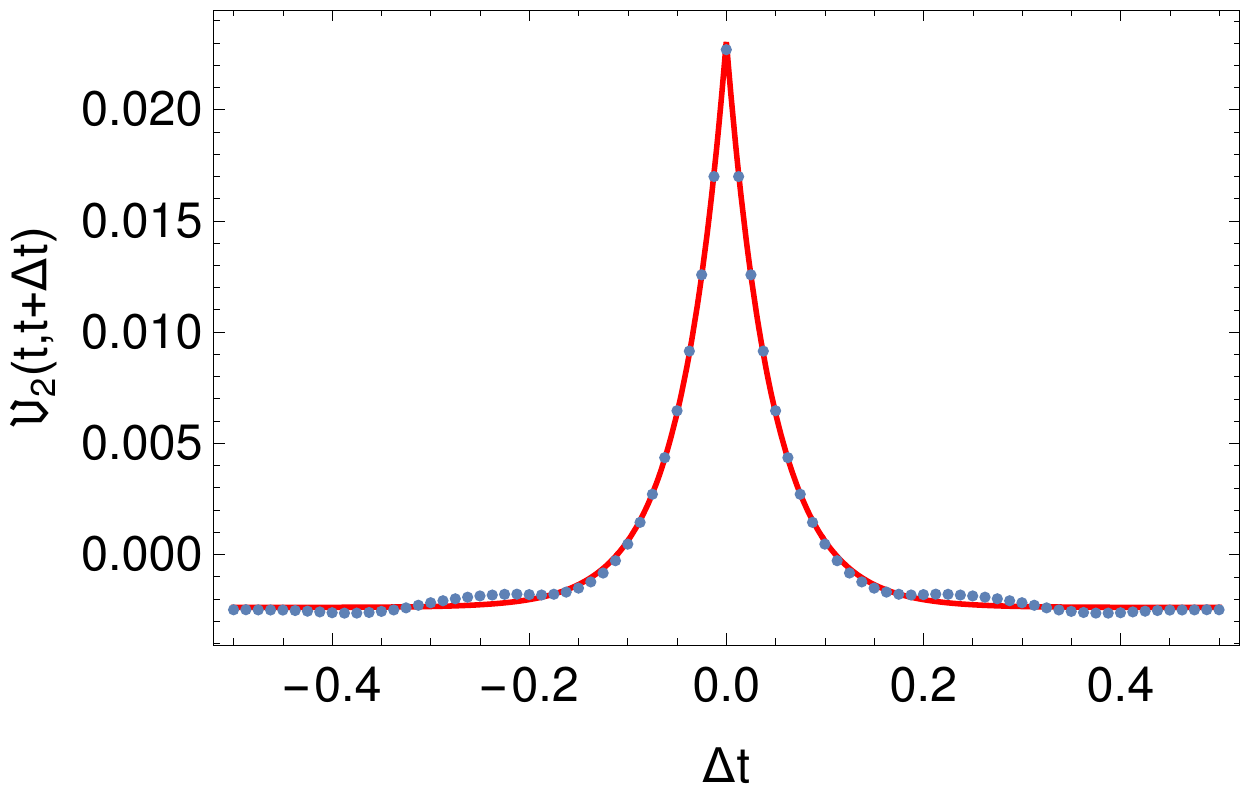}
\caption{The correlator $\mathfrak V_2(t,t+\Delta t)$ on the torus in arbitrary units. Blue dots are the lattice data (averaged over all temporal slices), the red line indicates the fitted analytical result.}
\label{fig:torus}
\end{figure}

We are now in the situation that we can compare to the data obtained from \CDT{} simulations on the torus for bare parameters {(cf.\ the discussion below eq.\ \eqref{eq4})}
\begin{equation}\label{eq:config}
\kappa_0=2.2, \quad \Delta=0.6 \, ,
\end{equation}
and configurations built from $N_4=160000$ simplices \cite{Ambjorn:2016fbd}.
This point is located well within the de Sitter phase of the \CDT{} phase diagram \cite{Ambjorn:2018qbf}. Averaging over all times, we can fit the correlator $\mathfrak V_2(t,t+\Delta t)$ to extract the value of Newton's constant $G_N$ and the mass parameter $b^2$. A least squares fit gives
\begin{equation}\label{Torusfit}
 G_N = 0.14 \, a_\text{CDT}^2 \, , \enspace \ell_\text{Pl} = 0.37 \, a_\text{CDT} , \enspace b = 6.93/a_\text{CDT} \, .
\end{equation}
Here $a_\text{CDT}$ is the lattice spacing and $\ell_\text{Pl} \equiv  \sqrt{G_N}$ is the Planck length. We display the lattice data (blue dots) and our fit (red line) in \autoref{fig:torus}, and find a very good agreement between the two.

The relation between the lattice spacing and the physical radius $r$ of the torus can be obtained by fitting the eigenvalues of the covariance matrix to the analytic form \eqref{eq:toruseigensol}. Since the higher eigenvalues are less precise, we only fit the lowest three eigenvalues. Demanding that the resulting Newton's constant agrees with the one from the fit, gives a relation between the physical radius and the lattice spacing
\begin{equation}
r = 3.09 \, a_\text{CDT} \, .
\end{equation}

The value of $b$ can also be calculated from the lattice data directly by {resorting to an action for the volume fluctuations inspired by a mini-superspace computation} 
\begin{equation}
 b_\text{CDT} = \sqrt{\frac{\bar\Gamma u}{\gamma(1+\gamma)}} \, \bar V_3 \, .
\end{equation}
Here, $\bar\Gamma$ is related to the kinetic term, $u$ to the potential term and $\gamma$ is a critical exponent. For the data point \eqref{eq:config}, we take $\bar\Gamma=26.3$, $u=1.30\times10^{-6}$ ("first average then invert"), $\bar V_3=2000$ and $\gamma=1.16$ given in \cite{Ambjorn:2016fbd}, and find
\begin{equation}
 b_\text{CDT} = (7.39 \pm 0.84 \pm 0.14 \Delta\bar\Gamma) / a_\text{CDT} \, ,
\end{equation}
which is consistent with the fit value within numerical precision. No numerical error $\Delta\bar\Gamma$ has been given in \cite{Ambjorn:2016fbd}, and we leave it unspecified here.

Naturally, one would expect that the local part of the quantum effective action also contains higher-derivative terms like $\int d^4x \sqrt{g}R^2$ containing four (or more) spacetime derivatives. On the toroidal background the Hessian $\Gamma^{(2)}$ then acquires an additional term proportional to $\partial_t^4$. Including this contribution in the fitting procedure shows that the related coefficient is negligible though.

Adapting the construction \eqref{eq11} to backgrounds where $\bar{\Sigma} = S^3$ leads to the eigenproblem studied in \cite{Ambjorn:2008wc}. The authors have shown there that the resulting fluctuation spectrum agrees very well with the numerical data. For the configuration \eqref{eq:config} a comparison of the lowest eigenvalue of the covariance matrix with the continuum version yields
\begin{equation}\label{spherefit}
G_N = 0.23 \, a_\text{CDT}^2 \, , \enspace \ell_\text{Pl}  = 0.48 \, a_\text{CDT} \, .
\end{equation} 
The relation between the lattice spacing and the physical radius  $r=3.1 \, a_\text{CDT}$ is taken from \cite{Ambjorn:2008wc} and agrees with the relation on the torus at the same bare parameters. Thus our continuum approach also reproduces the Monte Carlo results obtained for spherical topology.

Background independence of the quantum effective action suggests that the values for $G_N$ obtained in different simulations should agree.
The comparison of the Newton's constant found for the toroidal \eqref{Torusfit} and spherical topology \eqref{spherefit} indicates different values though.  This puzzle is resolved by the observation that the couplings $G_N$ obtained from the volume correlations on the torus and spherical background are actually not the same: for non-vanishing background curvature higher-order curvature terms contribute to the $\partial_t^2$-terms appearing in $\Gamma^{(2)}$, so that the $G_N$ obtained in \eqref{spherefit} is actually a function of the Newton's constant defined in \eqref{Glocal} and higher-derivative couplings.
Likewise, the fact that the data obtained for $\bar{\Sigma} = S^3$ suggest $b=0$ is not in contradiction with the toroidal results since the correlator of volume fluctuations evaluated on a background with non-zero curvature is not sensitive to this coupling. Instead it probes non-trivial endomorphism terms regulating the inverse Laplacians on a constant curvature background \cite{Barvinsky:2012mro}, which come with their own couplings.

\section{Information from complementary correlators}
The two-point autocorrelation function \eqref{volumecorrelator} gives access to some couplings appearing in the quantum effective action. A more complete picture can be developed by either studying correlation functions of different geometrical quantities or higher-order $n$-point functions.

\emph{Higher-order $n$-point functions --}
A natural generalisation of \eqref{volumecorrelator} are  
higher-order correlators involving the fluctuations of spatial volumes at $n$ time-steps
\be\label{Voln}
\mathfrak V_n(t_1,\ldots,t_n) \equiv \langle \delta V_3(t_1) \ldots \delta V_3(t_n) \rangle \, . 
\ee
These correlators can be constructed systematically by taking derivatives of $\mathfrak V_2$ w.r.t.\ a suitable source. The three-point correlator $\mathfrak V_3$ for instance then involves the three-point vertex 
contracted with three propagators. On a flat, toroidal background $\mathfrak V_n(t_1,\ldots,t_n)$ carries information on couplings associated with terms built from $n$ powers of the Riemann tensor (and its contractions). 

\noindent
\emph{$2$-point functions involving curvatures --} Complementary, one may study the autocorrelation of curvature fluctuations involving the extrinsic or intrinsic curvature. Focusing on one concrete example, we introduce the averaged intrinsic curvature
\begin{equation}
\mathcal R_3(t) = \int \text{d}^3x \sqrt{\sigma} \, {}^{(3)}R \, .
\end{equation}
The analogue of \eqref{volfluct} is then obtained by expanding $\mathcal R_3(t)$ in terms of fluctuations 
\be
\begin{split}
\tilde\sigma(x,t) = & a(t)\bar\sigma^{\mu\nu}(x)\delta\sigma_{\mu\nu}(x,t) \, , \\
\tilde\sigma_{\mu\nu}^\text{TL}(x,t) = & a(t) \delta\sigma_{\mu\nu}(x,t) - \tfrac{1}{3} \bar\sigma_{\mu\nu}(x)\,\tilde\sigma(x,t) \, .
\end{split}
\ee
Dropping integrals over total derivatives, fluctuations of the averaged intrinsic curvature are related to the fluctuation fields by
\begin{equation}\label{Rexp}
 \delta\mathcal R_3(t) = \int \text{d}^3x \sqrt{\bar\sigma} \left( \tfrac{1}{6} {{}^{(3)}\tilde R} \tilde\sigma - {{}^{(3)}\tilde S}^{\mu\nu} \tilde\sigma_{\mu\nu}^\text{TL} \right) \, ,
\end{equation}
where ${}^{(3)}\bar R = {{}^{(3)}\tilde R}/a(t)^2$ and ${}^{(3)}\bar S^{\mu\nu} = {{}^{(3)}\tilde S}^{\mu\nu}/a(t)^4$ indicate the background spatial Ricci scalar and trace-free spatial Ricci tensor, respectively. The autocorrelation function can then again be expressed in terms of the propagators of the fluctuation fields,
\begin{equation}\label{eq:curvcorr}
\begin{split}
 \langle \delta \mathcal R_3(t^\prime) \delta \mathcal R_3(t) & \rangle
 =  \, \int \text{d}^3 x \sqrt{\bar\sigma} \int \text{d}^3 y \sqrt{\bar\sigma} \Big[ \tfrac{{}^{(3)}\tilde R^2}{36} \mathfrak G_{\tilde\sigma\tilde\sigma} \\ & 
- \tfrac{1}{3} {}^{(3)}\tilde R \, {}^{(3)}\tilde S^{\mu\nu} \left(\mathfrak G_{\tilde\sigma\tilde\sigma^\text{TL}}\right)_{\mu\nu} \\&
 + {}^{(3)}\tilde S^{\mu\nu} \, {}^{(3)}\tilde S^{\rho\sigma} \left(\mathfrak G_{\tilde\sigma^\text{TL}\tilde\sigma^\text{TL}}\right)_{\mu\nu\rho\sigma} \Big] \, .
 \end{split}
\end{equation}
Notably $ \langle \delta \mathcal R_3(t^\prime) \delta \mathcal R_3(t) \rangle$ vanishes on a toroidal background since it is proportional to the background curvature. This feature is owed to terminating the expansion \eqref{Rexp} at leading order in the fluctuation fields. Once terms quadratic in the fluctuations are included, the correlator \eqref{eq:curvcorr} involves non-zero contributions related to the four-point vertex of the fluctuation fields.

\section{Conclusions}

We introduced a new research program to reverse-engineer the quantum effective action for  gravity from correlation functions. This provides for the first time a direct link between continuum and lattice in quantum gravity beyond abstract quantities like critical exponents \cite{Falls:2015cta, Biemans:2016rvp} and spectral dimensions \cite{Ambjorn:2005db, Lauscher:2005qz, Benedetti:2009ge, Clemente:2018czn}. Where lattice data was available, agreement with an Einstein-Hilbert action, potentially amended by a particular non-local interaction or higher-order scalar curvature terms, was found.

A particularly intriguing result is that the lattice simulations on the torus suggest the presence of a (non-local) mass term. The authors of \cite{Ambjorn:2016fbd} argue that the occurrence of this term is a genuine quantum gravity effect. Generically, it is expected that quantum fluctuations of massless particles induce these kind of non-local terms in the quantum effective action. A prototypical example is provided by quantum chromodynamics where such terms correctly describe the non-perturbative gluon propagator in the \IR{} \cite{Boucaud:2001st, Capri:2005dy, Dudal:2008sp, Pelaez:2014mxa}. 

In general, it is conceivable that non-local gravitational interactions provide a dynamical explanation of dark energy, without the need for a fine-tuned cosmological constant \cite{Maggiore:2013mea, Maggiore:2014sia, Foffa:2013vma}.
In particular, the non-local contribution {given in eq.\ \eqref{4a}}
%
%
%
forms a key part of the Maggiore-Mancarella cosmological model \cite{Maggiore:2014sia, Belgacem:2017cqo}, which has been highly successful in describing the cosmological evolution of the Universe.\footnote{For earlier work discussing the effect of non-local terms in cosmology also see \cite{Deser:2007jk}.}
It is intriguing that the non-classical behaviour seen on the lattice is compatible with such a non-local quantum effect. {Since \eqref{4b} is known to lead to unstable modes in the cosmic perturbation spectrum \cite{Cusin:2015rex}, it would be highly interesting to determine $\tilde{b}$ from the fundamental formulation. This is beyond the present work though.} 

In this work, we analysed the arguably simplest non-trivial correlation function describing the autocorrelation of 3-volume fluctuations at two different times \eqref{volumecorrelator}. The systematic extension to correlation functions of higher order or other structures is evident. Notably, a measurement of the correlators \eqref{Voln} and \eqref{eq:curvcorr} may be actually feasible within the \CDT{} program, thereby providing further information on the quantum effective action. In particular, correlation functions of the averaged intrinsic curvature may be obtained by summing deficit angles or new sophisticated methods to measure curvature on general quantum spacetimes recently introduced in \cite{Klitgaard:2017ebu, Klitgaard:2018snm}. We stress that the construction of reverse-engineering the quantum effective action from correlators is not limited to \CDT{}, but applies to any given theory of quantum gravity in which the corresponding quantities can be computed. 

\medskip

\acknowledgments

We thank J. Ambj\o{}rn, A. G\"orlich and R. Loll
for interesting discussions, and A. G\"orlich for providing us with the numerical data from the \CDT{} simulations.
Interesting discussions with J. Donoghue, N. Klitgaard and M. Maggiore are acknowledged.
This research is supported by the Netherlands Organisation for Scientific Research (NWO) within
the Foundation for Fundamental Research on Matter (FOM) grant
13VP12.

\bibliography{general_bib}

\end{document}